\documentstyle[epsf]{article}
\def\gtsim{\mathrel{\raise .5mm \hbox{$>$} 
\kern-2.7mm \lower 1mm \hbox{$\sim$}}}
\def\ltsim{\mathrel{\raise .5mm \hbox{$<$} 
\kern-2.7mm \lower 1mm \hbox{$\sim$}}}
\begin{document}
\title{
Large-$q$ expansion of the energy and magnetization cumulants \\
for the two-dimensional $q$-state Potts model}
\author{
        H. ARISUE \\
        Osaka Prefectural College of Technology, \\
        Saiwai-cho, Neyagawa, Osaka 572, Japan
              \and
        K. TABATA \\
        Osaka Institute of Technology, Junior College, \\
        Ohmiya, Asahi-ku, Osaka 535, Japan}
\maketitle
\begin{abstract}
We have calculated the large-$q$ expansion for the energy cumulants 
and the magnetization cumulants at the phase transition point 
in the two-dimensional $q$-state Potts 
model to the 21st or 23rd order in $1/\sqrt{q}$ using the finite lattice 
method. 
The obtained series 
allow us to give very precise estimates of the cumulants for $q>4$ 
on the  first order transition point. The result confirms us 
the correctness of 
the conjecture by Bhattacharya {\em et al.} on the asymptotic behavior 
not only of the energy cumulants but also the magnetization cumulants 
for $q \rightarrow 4_+$. 

\vspace{0.5cm}
\end{abstract}

\section{Introduction}
It is a challenging problem to determine the order 
of the phase transition in the physical system where the order 
changes from the first order to the second one by a small variation of 
a parameter of the system.
It is important to know how the quantities that remain finite 
at the first order transition point diverge or vanish 
when the parameter approaches the point at which the transition 
changes to the second order one.
The $q$-state Potts model\cite{Potts,Wu} in two dimensions gives
a good place to investigate the properties of the phase transition 
around such a point.
It exhibits the first order phase transition for $q>4$
and the second order one for $q\le 4$.
Many quantities of this model 
at the phase transition point $\beta=\beta_t$ are known exactly 
for $q>4$, including the 
latent heat\cite{Baxter1973} and the correlation 
length\cite{Klumper,Buffenoir,Borgs},
which vanishes and diverges, respectively, 
in the limit $q\rightarrow 4_+$. 
One the other hand physically important quantities 
such as the specific heat 
and the magnetic susceptibility at the transition point,
which also increase to infinity as $q\rightarrow 4_+$, 
are not solved exactly.
Bhattacharya {\em et al.}\cite{Bhattacharya1994} made 
a stimulating conjecture
on the asymptotic behavior of the energy cumulants 
including the specific heat
at the first order transition point; 
the relation between the energy cumulants and the correlation length
in their asymptotic behavior at the first order transition point 
for $q\rightarrow 4_+$ 
will be the same as their relation in the second order phase transition
for $q=4$ and $\beta\rightarrow \beta_t$, 
which is well known from their critical exponents.
If this conjecture is true not only in the $q$-state Potts model 
but also in other general physical systems with a phase transition 
whose order change from the first order to 
the second one when some parameter of the system is varied,
it would give a good guide in investigating such systems.

Bhattacharya {\em et al.}\cite{Bhattacharya1997} also
made the large-$q$ expansion of the energy cumulants 
to order 10 in $z\equiv 1/\sqrt{q}$
and they analyzed the series according to the conjecture, 
giving the estimates of the cumulants
at the first order transition point for $q \ge 7$. 
The obtained estimates are at least one order of magnitude better than 
those given by other methods 
such as the Monte Carlo simulations\cite{Janke} and the 
low-(and high-)temperature 
expansions\cite{Bhanot1993,Briggs,Arisue1997},
in spite of the fact that 
the maximum size of the diagrams taken into account 
in their large-$q$ expansion is less than the correlation length 
for $q\ltsim 15$ 
while the Monte Carlo simulations are done carefully 
on the lattice with its size about 30 times as large as 
the correlation length 
and also the fact that the maximum size of the diagrams 
in their large-$q$ 
expansion is as small as the half of the maximum size of the diagrams 
in the low-(and high-)temperature expansions
These strongly suggest the correctness of the conjecture. 
The large-$q$ series obtained by them are, however, 
not long enough to 
investigate the behavior of the energy cumulants for $q$ closer to $4$.

In this paper we will enlarge the large-$q$ series 
not only for the energy cumulants but also the magnetization cumulants 
at the transition point to order of 21 or 23 in $z$ 
using the finite lattice 
method\cite{Enting1977,Creutz,Arisue1984},
instead of the standard diagrammatic method 
used by Bhattacharya {\em et al.}. 
The finite lattice method can in general give longer series 
than those generated by the diagrammatic method 
especially in lower spatial dimensions. 
In the diagrammatic method, 
one has to list up all the relevant diagrams and count the number 
they appear.
In the finite lattice method we can skip this job and 
reduce the main work to the calculation of the expansion 
of the partition function for a series of finite size lattices,
which can be done using the straightforward site-by-site 
integration\cite{Enting1980,Bhanot1990} 
without the diagrammatic technique.
This method has been used mainly to 
generate the low- and high-temperature series in statistical systems 
and the strong coupling series in lattice gauge theory.
One of the purposes of this paper is to demonstrate that this method  
is also applicable to the series expansion 
with respect to the parameter other than the 
temperature or the coupling constant.
Such an approach was successfully done by Bakaev and Kabanovich 
in generation the long series for the ground-state entropy of the $q$-state
antiferromagnetic Potts model in two dimensions\cite{Bakaev1994}.
The long large-$q$ series obtained here by the finite lattice method 
enables us to examine the conjecture by Bhattacharya {\it et al.} on
the asymptotic behavior of the energy and magnetization cumulants for $q$ 
very close to 4. 
We can also give the estimates of these cumulants 
for each $q=5,6,7,\cdots$ that are orders of magnitude more precise 
than the estimates by the previous works and they would serve 
as a target in investigating this model in various contexts, 
for example in testing 
the efficiency of a new algorithm of the numerical simulation.
A brief note on this subject was already presented 
in \cite{Arisue1998a},
where we concentrated only on the specific heat.

The remainder of this paper is organized as follows.
The algorithm of the finite lattice method to generate 
the large-$q$ expansion 
of the free energy is given for the disordered phase in section 2 
and for the ordered phase in section 3. 
The relation between the two algorithms is also discussed in section 3.
The obtained series for the energy and magnetization cumulants 
at the phase transition points
are presented in section 4.
In section 5 the series are analyzed to investigate the 
asymptotic behavior of the cumulants at $q\rightarrow 4_+$ as well as 
to make the estimation of the cumulants at various values of $q>4$.
In section 6 we conclude with a brief summary and a few remarks.

\section{Algorithm in the disordered phase}
The model is defined by the partition function
\begin{equation}
  Z=\sum_{\{s_i\}} \exp{\left(\beta \sum_{\langle i,j \rangle}\delta_{s_i,s_j}
          +h\sum_{i}\delta_{s_i,1}
          \right)}\;,  \label{eq:partitionf}
\end{equation}
where the spin variable $s_i$ at each site $i$ 
takes the values $1,2,\cdots,q$
and $\langle i,j \rangle$ represents the pair of nearest neighbor sites.
The phase transition point $\beta_t$ for $h=0$ is given 
by $\exp{(\beta_t)-1=\sqrt{q}}$.
We consider below in this section 
the free energy density in the disordered phase, 
which is given by 
\begin{equation}
F_d(\beta) =
\lim_{L_x,L_y\rightarrow \infty}(L_x L_y)^{-1}\ln{Z_d(L_x,L_y)}\;. 
\label{eq:Free_energy}
\end{equation}
Here the partition function $Z_d(L_x,L_y)$ in the disordered phase 
should be calculated for the $L_x\times L_y$ lattice 
with the free boundary condition as depicted in Fig.1.
\begin{figure}[tb]
\epsfysize=4.8cm
\epsffile{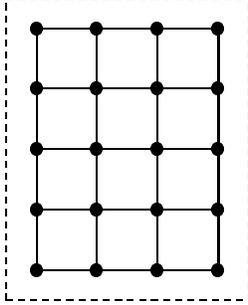}
\caption{
$L_x\times L_y$ lattice (in this figure $L_x=4$ and $L_y=5$)
with the free boundary condition for the disordered phase.
}
\label{fig1}
\end{figure}
We first restrict ourselves to the discussion in the case of $h=0$.
The large-$q$ expansion of the partition function in this phase
can be given through the 
Fortuin-Kasteleyn representation\cite{Fortuin} as
\begin{equation}
   Z_d(L_x,L_y)
  = q^{L_x L_y}\left[1+\sum_{X\subseteq \Lambda_d} \phi(X)  \right]\;, 
     \label{eq:partitionfFK}\\
\end{equation}
with
\begin{eqnarray}
\phi(X) &=& \left(e^\beta -1\right)^{b(X)}q^{c(X)-L_x L_y}\nonumber\\
        &=& Y^{b(X)}z^{2[L_x L_y-c(X)]-b(X)}\;, \label{eq:phi_d}
\end{eqnarray}
Here $\Lambda_d$ is the set consisting of all bonds in the lattice, 
$X$ is any subset of $\Lambda_d$ and represents 
a configuration of bonds, 
excluding the empty set (the null configuration) 
which gives the leading contribution and 
is separated in the first term of Eq.(\ref{eq:partitionf}). 
In Eq.(\ref{eq:phi_d}) $Y\equiv (e^{\beta}-1)/\sqrt{q}$, 
$z\equiv 1/\sqrt{q}$,
$b(X)$ is the number of bonds in $X$ and 
$c(X)$ is the number of clusters of sites in $X$. 
Two sites connected to each other by a bond belong to the same
cluster (An isolated single site is a cluster). 
The $\phi(X)$ is called the {\it activity} of $X$.
It is useful to note that the activity can be rewritten as
\begin{eqnarray}
\phi(X) &=& \left(e^\beta -1\right)^{b(X)}q^{-b(X)+l(X)} \nonumber\\
        &=& Y^{b(X)}z^{b(X)-2l(X)}\;, \label{eq:phi_b}
\end{eqnarray}
where $l(X)$ is the number of independent closed loops formed 
by the bonds in $X$.

We define $H_d(l_x,l_y)$
for each $l_x \times l_y$ lattice ($1\le l_x\le L_x, 1\le l_y\le L_y$) 
as
\begin{equation}
   H_d(l_x,l_y) = \ln{ \left[Z_d(l_x,l_y)/q^{l_x l_y}\right]}\;, 
   \label{eq:H}
\end{equation}
where $Z_d(l_x,l_y)$ is the partition function 
for the $l_x \times l_y$ 
lattice, again with the free boundary condition as shown in Fig.1. 
We next define $W_d(l_x,l_y)$ recursively as
\begin{eqnarray}
&&   W_d(l_x,l_y) = H_d(l_x,l_y) \nonumber\\
&& \qquad - \sum_{\scriptstyle l_x^{\prime}\le l_x,l_y^{\prime}\le l_y
\atop\scriptstyle  (l_x^{\prime},l_y^{\prime})\ne (l_x,l_y)}
(l_x-l_x^{\prime}+1)(l_y-l_y^{\prime}+1)W_d(l_x^{\prime},l_y^{\prime})
\;. \label{eq:W}
\end{eqnarray}
Here the term with $l_x^{\prime}=l_x$ and $l_y^{\prime}=l_y$ should be 
excluded in the summation.
Then the free energy density defined by Eq.(\ref{eq:Free_energy}) 
is given 
by\cite{Arisue1984} 
\begin{equation}
    F_d(\beta) = \ln{(q)}+\sum_{l_x,l_y} W_d(l_x,l_y)\;. \label{eq:F}
\end{equation}

 Now we have to know from what order in $z$ the $W_d(l_x,l_y)$ starts.
Only for this purpose, 
let us suppose to apply the standard cluster 
expansion\cite{Domb1974,Muenster1981} based on the diagrams
the large-$q$ expansion for the free energy in the disordered phase. 
It should be stressed that in the practical calculation 
of the finite lattice 
method we need not perform this cluster expansion at all.
A {\it polymer} is defined as a set $X$ of bonds 
that satisfies the condition
that there is no division of $X$ into two sets $X_1$ and $X_2$ 
so that $\phi(X)=\phi(X_1)\phi(X_2)$.
A set $X$ composed of a single bond is a polymer.
Below in this paragraph we discuss on the polymers that consists of 
two or more bonds. 
Now we define a {\it cluster of bonds} in $X$ so that 
two bonds that are connected by a site belong to the same cluster.
A polymer consisting of two or more bonds is composed of 
a single cluster of bonds,
since a set of bonds composed of more than one  
disconnected clusters of bonds apparently does 
not satisfy the condition of a polymer. 
A cluster of bonds is a polymer 
if and only if the cluster of the bonds cannot be divided 
into tow or more clusters of bonds by cutting at any site 
that is shared by the bonds belonging to the cluster.
We call such a cluster of bonds as {\it one-site irreducible}.
In fact for a set $X_r$ consisting of a {\it one-site reducible} 
cluster of bonds that can be divided into clusters 
$X_1,X_2,\cdots,X_n$ of bonds by cutting at a site, we have 
$\phi(X_r)=\phi(X_1)\phi(X_2)\cdots\phi(X_n)$, which implies that 
this set of bonds is not a polymer.
For a cluster of bonds to be one-site irreducible,
it is necessary that the bonds should form one or more closed loops 
and that each bond should belong to at least one of the loops,
which we call {\it closed}.
An example of a closed cluster of bonds is shown in Fig.2(a).
For each closed cluster of bonds, 
the loop of the bonds that forms the perimeter of the whole cluster 
is called the {\it exterior boundary}. The exterior boundary for the 
closed cluster of bonds in  Fig.2(a) is indicated in Fig.2(b).
\begin{figure}[tb]
\epsfysize=4cm
\epsffile{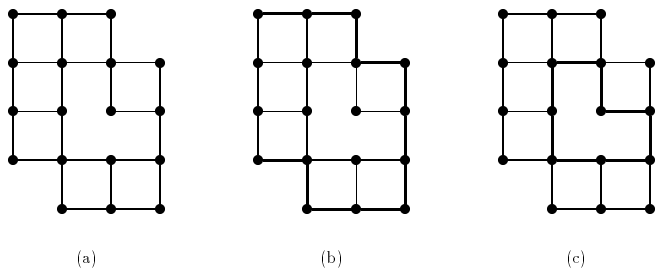}
\caption{
(a) Example of a closed cluster of bonds with a hole. \ \ 
(b) The same as (a) with the exterior boundary 
highlighted by thick lines. \ \ 
(c) The same as (a) with the interior boundary 
highlighted by thick lines.
}
\label{fig2}
\end{figure}
An {\it interior boundary} is a loop inside which all bonds are removed
and we call the part of the cluster 
surrounded by the interior boundary as a {\it hole}.
The closed cluster of bonds in  Fig.2(a) has a hole with the interior 
boundary, as indicated in Fig.2(c). 
For a cluster of bonds to be one-site irreducible,
not only should it be closed but also its exterior boundary
has no crossing.
If the exterior boundary has a crossing, as in Fig.3,
the cluster of bonds is one-site reducible and is not a polymer.
\begin{figure}[tb]
\epsfysize=4cm
\epsffile{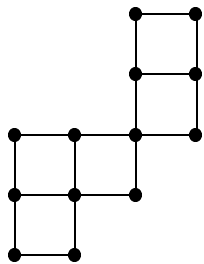}
\caption{
Example of a polymer whose exterior boundary has a crossing.
}
\label{fig3}
\end{figure}
Thus a polymer is the closed cluster of bonds 
with its exterior boundary having no crossing.
We list as an illustration in Fig.4 all the polymers that can be 
embedded within $3\times 3$ lattice with their activities. 
The polymers are skipped 
that can be obtained by rotating and/or reflecting those in Fig.4.
\begin{figure}[tb]
\epsfysize=8cm
\epsffile{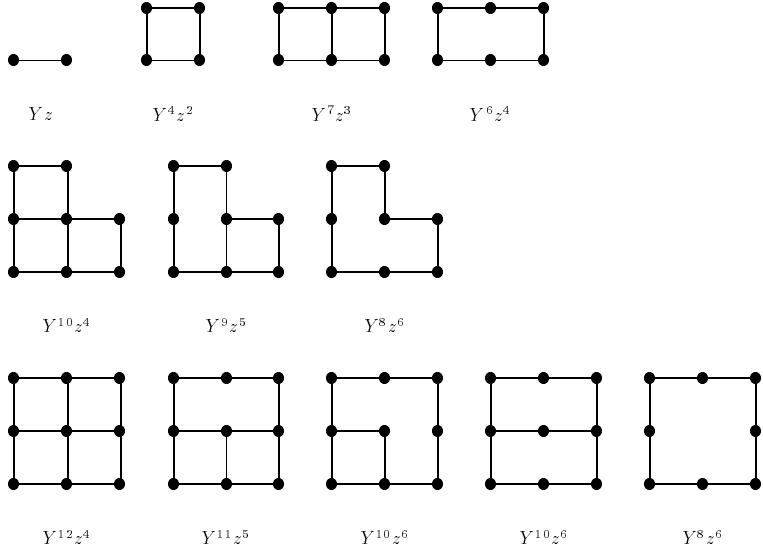}
\caption{
All the polymers that can be embedded within $3\times 3$ lattice 
with their activities. 
}
\label{fig4}
\end{figure}
For later convenience we add some definitions concerning a polymer.
A {\it concave corner} in the boundary of a polymer is the part of the 
boundary composed of two neighboring bonds that are 
bent perpendicular to each other toward the interior of the polymer.
If two neighboring corners on the boundary are both concave,
we call the two corners together with the bonds in a straight line 
connecting the two corners as a {\it concave part} of the boundary. 
Examples of the concave parts are given in Fig.5. 
\begin{figure}[tb]
\epsfysize=4cm
\epsffile{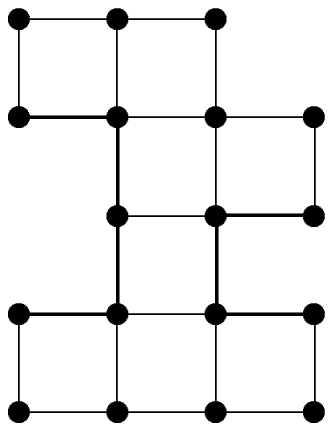}
\caption{
Example of a polymer with two concave parts highlighted by thick lines.
}
\label{fig5}
\end{figure}

According to the theorem of the standard cluster 
expansion\cite{Domb1974,Muenster1981}, 
the free energy for the $l_x\times l_y$ lattice is given by 
\begin{eqnarray}
&&   \ln{Z_d(l_x,l_y)} = l_xl_y\ln{q}           \nonumber\\
&&    +\sum_{\{X_1,X_2,\cdots,X_n:connected\}} 
        c(X_1,X_2,\cdots,X_n)\phi(X_1)\phi(X_2)\cdots\phi(X_n), 
        \label{eq:clusterx}
\end{eqnarray}
where the summation should be taken for every cluster 
of {\it connected} 
polymers $\{X_1,X_2,\cdots,X_n\}$ defined by the condition
that there is no division of the cluster of polymers 
$\{X_1,X_2,\cdots,X_n\}$ into two clusters of polymers 
$\{X_1^{\prime},X_2^{\prime},\cdots,X_m^{\prime}\}$ and
$\{X_1^{\prime\prime},X_2^{\prime\prime},\cdots,
X_{n-m}^{\prime\prime}\}$ 
($m<n$) so that 
$\phi(X_1\cup X_2\cup\;\cdots\;\cup X_n)
=\phi(X_1^{\prime}\cup X_2^{\prime}\cup\;\cdots\;\cup X_m^{\prime})
\phi(X_1^{\prime\prime}\cup X_2^{\prime\prime}\cup\;\cdots\;
\cup X_{n-m}^{\prime\prime})$.
(The `cluster' of polymers here should be distinguished from the 
`cluster' of sites in the Fortuin-Kasteleyn representation
or the `cluster' of bonds defined in the explanation of the polymers 
above.)
The $c(X_1,X_2,\cdots,X_n)$ in Eq.(\ref{eq:clusterx}) 
is the numerical factor that depends on the way how the clusters 
of polymers are connected, the details of which are not necessary here.

When a cluster of connected polymers can be embedded 
into the $l_x \times l_y$ lattice 
but cannot be embedded into any of its rectangular sub-lattices,
we call it shortly to {\it fit into} the $l_x \times l_y$ lattice. 
Then we can prove the theorem\cite{Arisue1984} 
that the Taylor expansion of the $W_d(l_x,l_y)$ with respect 
to $z$ includes the contribution from all the clusters of polymers 
in the standard cluster expansion 
that fit into the $l_x \times l_y$ lattice.

Now, according to this theorem, 
we can know the order of $W_d(l_x,l_y)$ in $z$ 
by examining the clusters of connected polymers 
that fit into the $l_x \times l_y$ lattice.
We first note that $W_d(1,1)=0$ since the $1\times 1$ lattice consists 
of a single site with no bond, 
so that no polymer can be embedded into this lattice,
and $W_d(2,1)=W_d(1,2)=O(z)$ since the $2\times 1$ or $1\times 2$ 
lattice consists of two sites connected by one bond, 
so that a polymer composed 
of a single bond can be embedded into this lattice. 
We also note that 
$W_d(l_x,1)=W_d(1,l_y)=0$ for $l_x\ge 3$ or $l_y\ge 3$, 
because the corresponding lattice consists of the sites connected 
by bonds all on a straight line and 
the only clusters of polymers 
that fit into the lattice are those with each polymer composed 
only of a single bond and these polymers are not connected
according to the definition of the connected polymers.

We next consider the order of $W_d(l_x,l_y)$ for $l_x\ge 2$ 
and $l_y\ge 2$.
We first consider a cluster consisting of a single polymer
that fits into the $l_x \times l_y$ lattice. 
If the polymer has a concave part in its exterior boundary 
we can add some bonds to make the part flat, 
obtaining a new polymer whose activity is changed 
by a factor of $Y^nz^{-1}$ 
with $n$ the number of added bonds, so its order in $z$ is lowered.
This implies that a polymer that has concave part 
in the exterior boundary
does not contribute to the lowest order term of $W_d(l_x,l_y)$ in $z$. 
Let us next consider the case when the polymer has some holes.
The interior boundary surrounding each hole has
at least one concave part. 
(We take the hole as the exterior of the polymer 
when we say `concave' here.)
We can add some bonds to make the concave part flat, 
again obtaining a new polymer with the order of its activity in $z$
lowered by a factor of $z^{-1}$. 
This implies that a polymer that has a hole 
does not contribute to the lowest order term of $W_d(l_x,l_y)$ in $z$.
Thus the remaining polymers that have neither a concave part on the 
exterior boundary nor a hole in the interior give the contribution 
to the lowest order term of $W_d(l_x,l_y)$,
examples of which are shown in Fig.6.
\begin{figure}[tb]
\epsfysize=4.8cm
\epsffile{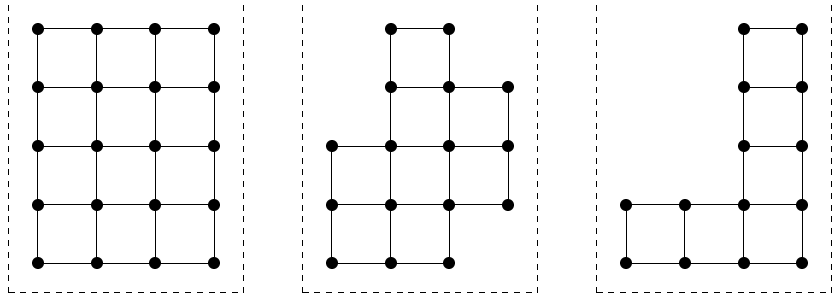}
\caption{
Examples of the cluster consisting of a single polymer 
that contribute to the lowest order term of the $W_d(l_x,l_y)$ 
with $l_x=4$ and $l_y=5$.
}
\label{fig6}
\end{figure}
They can be converted into each other by adding or removing pairs 
of bonds
to their concave corners or from their convex corners, respectively. 
When each pair of bonds is added or removed the activity 
of the polymer is changed by $Y^2$ with its order in $z$ unchanged.
Hence we see that their orders are all the same; 
that is, the order of $z^{l_x+l_y-2}$.
Next we consider a cluster of connected tow or more polymers 
that fits into the lattice.
It is enough to take into account the clusters of polymers that 
do not have either a concave part on the boundary or a hole 
in the interior,
since we are going to know the clusters of connected polymers 
that contributes to the lowest order term in $z$.  
Among the clusters of connected two or three polymers that fit into
the lattice, those give the lowest order contribution in which 
the polymers touch each other only through their exterior boundary and 
each neighboring pair of polymers share only one bond, like Fig.7(a),
and this type of clusters have the order of $z^{l_x+l_y+n-3}$ 
for the cluster of $n$ polymers ($n=2,3$), which is
higher than the lowest order contribution 
from the clusters composed of a single polymer.
\begin{figure}[tb]
\epsfysize=4.8cm
\epsffile{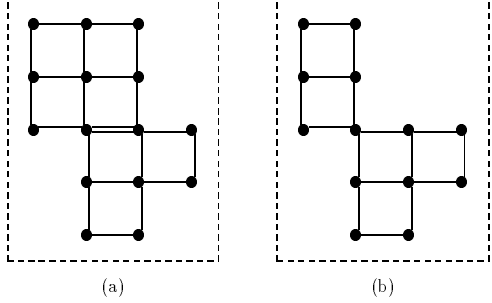}
\caption{
(a) Example of the cluster consisting of connected two polymers 
that share only one bond.
(b) Example of the cluster consisting of disconnected two polymers
that share only one site. 
}
\label{fig7}
\end{figure}
We note that 
a cluster of two or three polymers 
in which each neighboring pair of polymers 
share only one site, an example of which is 
given in Fig.7(b), is not connected 
according to the definition of the connected polymers.
Among the cluster of connected more than four or more polymers that 
fit into the lattice, those give the lowest order contribution 
in which four of the polymers surround a plaquette
and each neighboring pair of polymers share only one site,
like Fig.8, and this type of clusters have the order 
of $z^{l_x+l_y+n-4}$ 
($n\ge 4$) for the cluster of $n$ polymers, which is also
higher than the lowest order contribution 
from the clusters composed of a single polymer.
\begin{figure}[tb]
\epsfysize=4.8cm
\epsffile{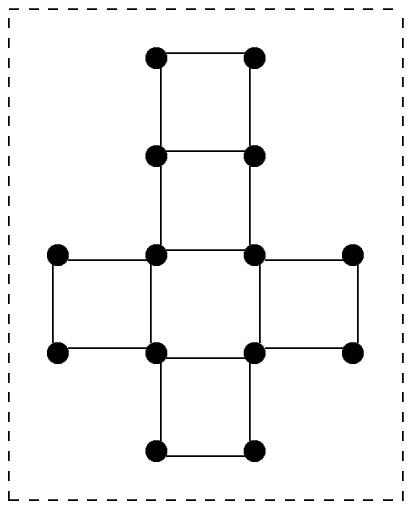}
\caption{
Example of the cluster consisting of connected four polymers 
that surround a plaquette 
and each neighboring pair of polymers share only one site.
}
\label{fig8}
\end{figure}
In summary the lowest order term of $W_d(l_x,l_y)$ has the order of 
$z^{l_x+l_y-2}$ for $l_x\ge 2$ and $l_y\ge 2$, 
which comes from the clusters of a single polymer 
without a hole or a concave part fitting into $l_x\times l_y$ lattice. 
Thus in order to obtain the series to order $z^N$ for the free energy 
density, we have only to take into account in Eq.(\ref{eq:F})
all the rectangular lattices that satisfy $l_x+l_y-2 \le N$.

We summarize the procedure of the finite lattice method to generate 
the large-$q$ expansion series of the free energy density 
in the disordered phase.\\
1) Calculate the partition function $Z_d(l_x,L_y)$ defined by 
Eq.(\ref{eq:partitionfFK}) as a power series of $z$
to order $z^N$ (and of $Y$ in full order) 
for each $l_x\times l_y$ lattice
with $l_x+l_y-2 \le N$. In these calculations use the transfer matrix 
method and integrate bond by bond\cite{Enting1980,Bhanot1990}, 
which reduces the 
necessary CPU time and storage.\\
2) Calculate $H_d(l_x,l_y)$ defined by Eq.(\ref{eq:H}) 
as a power series 
of $z$ by Taylor expanding $\ln{Z_d(l_x,l_y)}$
for each $l_x\times l_y$ lattice.\\
3) Calculate $W_d(l_x,l_y)$ defined recursively by Eq.(\ref{eq:W})
for each $l_x\times l_y$ lattice.\\
4) Sum up all the $W_d(l_x,l_y)$'s with $l_x+l_y-2 \le N$ 
according to Eq.(\ref{eq:F}) 
to obtain the expansion series for the free energy density.

Now we discuss on the free energy in the case of $h\ne 0$.
The Boltzman factor in Eq.(\ref{eq:partitionf}) 
that involves the effect of the magnetic field $h$ can be divided 
into two parts as
\begin{equation}
   \exp{\left(h\delta_{s_i,1} \right)}
        = 1+\mu\delta_{s_i,1}\;  \label{eq:Boltzmanh}
\end{equation}
with $\mu=e^{h}-1$.
The partition function $Z_d(l_x,l_y)$ is given  by the same expression as 
Eq.(\ref{eq:partitionfFK}) with $X$ now representing the configuration 
of both bonds and {\em frozen} sites. 
These frozen sites in $X$ are the sites whose spin variables 
are fixed to $1$ corresponding to the second term of
Eq.(\ref{eq:Boltzmanh}). 
The activity $\phi(X)$ is now 
given by the right hand side of Eq.(\ref{eq:phi_d}) 
multiplied by the factor $\mu^{s(X)}$ with $s(X)$ representing the number of
the frozen sites in $X$ and by the factor $q^{-f(X)}=z^{2f(X)}$ with $f(X)$ 
representing the number of the clusters of sites 
that include at least one frozen site in $X$.
$H_d(l_x,l_y)$ and $W_d(l_x,l_y)$ are defined by the same equation 
as Eq.(\ref{eq:H}) and (\ref{eq:W}) 
and the free energy density in the thermodynamic limit is given by 
Eq.(\ref{eq:F}). 
In the standard cluster expansion 
the polymer whose activity is proportional to $\mu^n$
is given by the cluster of bonds and $n$ frozen sites 
that cannot be divided into two or more clusters by cutting at a site 
so that one of the clusters involves no frozen site.
There is only one polymer 
whose activity is proportional to $\mu$, 
which is composed of a single frozen site with no bond. 
Therefore, if we expand $W_d(l_x,l_y)$ in $\mu$ as
$$W_d(l_x,l_y)=\sum_{n=0}^{\infty} W_d^{(n)}(l_x,l_y)\ \mu^n\;,$$
we know that 
$W_d^{(1)}(l_x,l_y)=z^2$ for $l_x=l_y=1$ and 
$W_d^{(1)}(l_x,l_y)=0$ for the other pairs of $l_x$ and $l_y$.
So the coefficient of the term proportional to $\mu$ 
in the free energy density is just $z^2$.
This will lead to the result that the magnetization in the disordered phase
is zero. 
We can see for $n\ge 2$ that $W_d^{(n)}(l_x,l_y)$ 
has the order of $z^{(l_x+l_y)}$ for $n\le l_xl_y$ 
and has the order of $z^{k(l_x+l_y)}$ or higher 
with $k=[n/(l_xl_y)]$ for $n>l_xl_y$. 
Here $[p]$ indicates the integer that is less than $p$.
The cluster of polymers that contributes to its lowest order term 
is composed of a single polymer that has $n$ frozen sites 
and fits into $l_x\times l_y$ lattice for $n\le l_xl_y$ 
and it is composed of $k$ polymers ($k\ge 1$) 
each of which has $l_xl_y$ frozen sites and fits into $l_x\times l_y$ lattice 
and a polymer with $n-kl_xl_y$ frozen sites 
for $n>l_xl_y$.
Thus the order of $W_d^{(n)}(l_x,l_y)$ for $n\ge 2$ is 
$z^{l_x+l_y}$ or higher. 
So, in order to generate the large-$q$ expansion of the term in 
the free energy density that is proportional to $\mu^n$ with $n\ge 2$,
it is enough to take into account the finite-size lattices that 
satisfy $l_x+l_y\le N$ in Eq.(\ref{eq:F}).

\section{Algorithm in the ordered phase}
Next we consider the large-$q$ expansion of 
the free energy density $F_o(\beta)$ in the ordered phase,
which is given by the same equation as 
Eq.(\ref{eq:Free_energy}) with the suffix $d$ replaced by $o$ where
the partition function $Z_o(L_x,L_y)$ in the ordered phase should be 
calculated for the $L_x\times L_y$ lattice 
with the fixed boundary condition as depicted in Fig.9.
The values of the spins on the boundary are all fixed to a single 
value among $1,2,\cdots,q$. Here we take this value as $1$.
\begin{figure}[tb]
\epsfysize=4.8cm
\epsffile{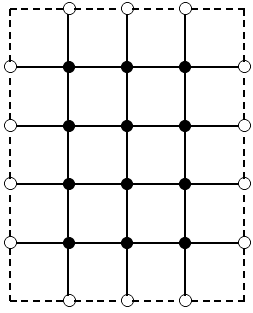}
\caption{
$L_x\times L_y$ lattice (in this figure $L_x=4$ and $L_y=5$)
with the fixed boundary condition for the ordered phase. 
The spins of the sites on the boundary (open circles) are fixed 
to a single value among $1,2,\cdots,q$.
}
\label{fig9}
\end{figure}

We first restrict ourselves to the discussion in the case of $h=0$ again.
By the duality the ordered and disordered free energy densities 
are related as
\begin{eqnarray}
 F_d(\beta)=F_o(\tilde{\beta})+2\ln((e^{\tilde{\beta}}-1)/\sqrt{q})
    \nonumber\\
     \mbox{for} \qquad (e^{\beta}-1)(e^{\tilde{\beta}}-1) = q\;.
       \label{eq:duality}
\end{eqnarray}
One method to obtain the large-$q$ series 
for the energy cumulant $F_o^{(n)}$ 
at $\beta=\beta_t$ in the ordered phase is to use the relations
\begin{eqnarray}
    F_d^{(1)}+F_o^{(1)}&=&  2 (1+z)\;,\nonumber\\
    F_d^{(2)}-F_o^{(2)}&=& -z [F_d^{(1)}-F_o^{(1)}]\;,\nonumber\\
    F_d^{(3)}+F_o^{(3)}&=& -3z [F_d^{(2)}+F_o^{(2)}]
                       +(z-z^2) [F_d^{(1)}+F_o^{(1)}]\;, \nonumber\\
    F_d^{(4)}-F_o^{(4)}&=& -6z [F_d^{(3)}-F_o^{(3)}]
                       -(z-6z^3) [F_d^{(1)}-F_o^{(1)}]\;, \nonumber\\
    F_d^{(5)}+F_o^{(5)}&=& -10z [F_d^{(4)}+F_o^{(4)}]
                       -(5z-60z^3) [F_d^{(2)}+F_o^{(2)}] \nonumber\\
           && +(z-z^2-24z^3+24z^4)[F_d^{(1)}+F_o^{(1)}]\;, \nonumber\\
    F_d^{(6)}-F_o^{(6)}&=& -15z [F_d^{(5)}-F_o^{(5)}]
                     -(15z-300z^3) [F_d^{(3)}-F_o^{(3)}] \nonumber\\
             &&-(z-90z^3+360z^5) [F_d^{(1)}-F_o^{(1)}]\;, \nonumber\\
       \label{eq:Duality_cumu}
\end{eqnarray}
which are derived directly 
from the duality relation (\ref{eq:duality}). 
These were given in Ref.\cite{Bhattacharya1997} 
except for the last one.

In spite of that,
we give below the algorithm to generate the series 
for the free energy density in the ordered phase 
by the finite lattice method without relying on the duality relation. 
This algorithm will prove important 
when we extend our work to generate the magnetization cumulants,
since they can be obtained by the derivative of the free energy 
density with respect to the external magnetic field and 
we have no longer duality relation for the free energy density 
in the presence of the external magnetic field. 
The large-$q$ expansion of the partition function in this phase 
can be given also through the 
Fortuin-Kasteleyn representation as
\begin{equation}
   Z_o(L_x,L_y)
     = \left(e^\beta -1\right)^{2L_x L_y-L_x-L_y}\left[1+
     \sum_{\tilde{X}} \varphi(\tilde{X})\right]\;, 
     \label{eq:partitionf_o}\\
\end{equation}
with
\begin{eqnarray}
\varphi(\tilde{X}) &=& 
\left(e^\beta -1\right)^{-b(\tilde{X})}q^{c(\tilde{X})}\nonumber\\
     &=& \tilde{Y}^{b(\tilde{X})}z^{b(\tilde{X})-2c(\tilde{X})}\;. 
         \label{eq:phi_o}
\end{eqnarray}
Here $\tilde{X}$ is a configuration of the removed bonds 
and we also use $\tilde{X}$  to represent the set of removed bonds.
In the leading configuration for the fixed boundary condition 
all the pairs of nearest neighbor sites are connected by the bonds
and there is no removed bonds ($\tilde{X}$ is the empty set),
and its contribution is separated as 
the first term of Eq.(\ref{eq:partitionf_o}).
In Eq.(\ref{eq:phi_o})  $\tilde{Y}\equiv \sqrt{q}/(e^{\beta}-1)$ and
$b(\tilde{X})$ is the number of bonds in $\tilde{X}$ and 
$c(\tilde{X})$ is the number of clusters of sites in the configuration 
of bonds with $\tilde{X}$ removed
without taking into account the clusters including the sites 
on the boundary.
We define $H_o(l_x,l_y)$ for each $l_x \times l_y$ lattice 
($1\le l_x\le L_x, 1\le l_y\le L_y$) as
\begin{equation}
   H_o(l_x,l_y) = \ln{ \left[Z_o(l_x,l_y)/
  \left(e^\beta -1\right)^{2l_x l_y-l_x-l_y}\right]}\;, \label{eq:H_o}
\end{equation}
where $Z_o(l_x,l_y)$ is the partition function 
for the $l_x \times l_y$ lattice, 
again with the fixed boundary condition as shown in Fig.9. 
We next define $W_o(l_x,l_y)$ recursively in the same way as 
in Eq.(\ref{eq:W}).
Then the free energy density in the ordered phase is given by
\begin{equation}
    F_o(\beta) = 2\ln{\left(e^\beta -1\right)}+\sum_{l_x,l_y} 
           W_o(l_x,l_y)\;.     \label{eq:F_o}
\end{equation}

It is useful to compare now the large-$q$ expansions by the finite 
lattice method for the disordered phase presented 
in the previous section and for the ordered phase presented here.
We first note that
the lattice with the free boundary condition in Fig.1
and the lattice with the fixed boundary condition in Fig.9 are
dual to each other 
in the sense that each site of one lattice is located 
at the center of a plaquette of the other lattice, 
when the two lattices have the same size. 
We denote $X\sim \tilde{X}$ as the relation 
between the configuration $X$
of the bonds on the lattice with the free boundary condition 
and the configuration $\tilde{X}$ of the removed bonds 
on the lattice with the fixed boundary condition, 
if the location of each bond (identified with its center) in $X$ 
coincides with the location of each corresponding removed bond 
(also identified with its center) in $\tilde{X}$ 
(see Fig.10(a) and (b)).
\begin{figure}[tb]
\epsfysize=4.8cm
\epsffile{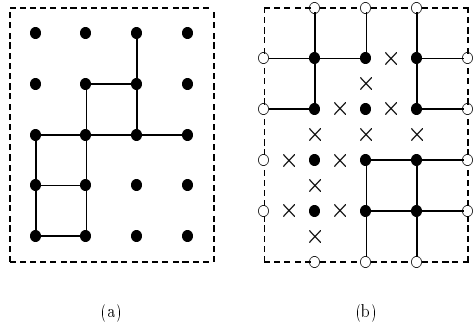}
\caption{
(a) Example of the configuration $X$ of bonds (solid lines) 
for the free boundary condition. 
(b) The configuration $\tilde{X}$ of removed bonds (crosses)
for the fixed boundary condition with $X\sim \tilde{X}$. 
}
\label{fig10}
\end{figure}
Then, the number $l(X)$ of independent closed loops for 
each configuration $X$ in Eq.(\ref{eq:phi_b}) 
is equal to the number $c(\tilde{X})$ of clusters of sites 
in the configuration with $\tilde{X}$ removed 
in (\ref{eq:phi_o}) for $\tilde{X}\sim X$ (see Fig.10(a) and (b)), 
so that $\phi(X)=\varphi(\tilde{X})$ when $Y=\tilde{Y}$. 
We therefore know from Eq.(\ref{eq:partitionfFK}), (\ref{eq:H}), 
(\ref{eq:partitionf_o}) and (\ref{eq:H_o}) that
$H_d(l_x,l_y;Y,z)=H_o(l_x,l_y;\tilde{Y},z)$ when $Y=\tilde{Y}$, 
hence the same is true for $W_d(l_x,l_y;Y,z)$ 
and $W_o(l_x,l_y;\tilde{Y},z)$.
Thus $W_o(l_x,l_y)$ has the same order in $z$ as $W_d(l_x,l_y)$, 
that is, the order of $z^{l_x+l_y-2}$. 
So in order to generate the series to order $z^N$ 
we have only to take into account the rectangular lattices with
$l_x+l_y-2\le N$ in Eq.(\ref{eq:F_o}).
This conclusion can also be reached by repeating the discussion 
based on the standard cluster expansion using the diagrams.
The discussion goes completely in parallel with those given 
for the disordered phase in the previous section 
if the set of bonds $X$ representing the configuration of bonds
for the disordered phase 
is replaced by the set of removed bonds $\tilde{X}$
for the ordered phase.

The relation of $W_d(l_x,l_y;Y,z)=W_o(l_x,l_y;\tilde{Y},z)$ 
for $Y=\tilde{Y}$ 
also implies that the free energy densities 
$F_d(\beta)$ given by Eq.(\ref{eq:F}) and 
$F_o(\tilde{\beta})$ given by Eq.(\ref{eq:F_o}) 
satisfy the duality relation (\ref{eq:duality}),
because the condition that $Y=\tilde{Y}$ is equivalent 
to the condition for $\beta$ and $\tilde{\beta}$ 
in the duality relation in Eq.(\ref{eq:duality}).

Now we proceed to the discussion of the free energy in the case of $h\ne 0$.
The partition function $Z_o(l_x,l_y)$ is now given  by the same expression as 
Eq.(\ref{eq:partitionf_o}) with $\tilde{X}$ representing the configuration 
of both removed bonds and frozen sites. 
The activity $\phi(\tilde{X})$ here 
is given by the right hand side of Eq.(\ref{eq:phi_o}) 
multiplied by the factor $\mu^{s(X)}$ with $s(X)$ 
representing now the number of 
the frozen sites in $\tilde{X}$ and 
by the factor $q^{-f(\tilde{X})}=z^{2f(\tilde{X})}$ with $f(\tilde{X})$ 
representing the number of the clusters of sites 
that include at least one frozen site in $\tilde{X}$
without taking into account the clusters including the sites on the boundary.
In the standard cluster expansion 
the polymers whose activity is proportional to $\mu^n (n\ge 1)$
are given by the same polymers 
that appear in the cluster expansion 
in the case of $h=0$ discussed above 
except that their $n$ sites are frozen. 
These frozen sites should belong to the cluster of 
sites that does not include the site on the boundary. 
By freezing $n$ sites the order of its activity in $z$ 
increases by $2n$.
There is one exceptional polymer. It is composed only of a single frozen site
with no removed bonds and its activity is just $\mu$. 
If we expand $W_o(l_x,l_y)$ in $\mu$ as
$$W_o(l_x,l_y)=\sum_{n=0}^{\infty} W_o^{(n)}(l_x,l_y)\ \mu^n\;,$$
we know that 
$W_o^{(n)}(l_x,l_y)=0$ for $l_x=1$ or $l_y=1$ since such a lattice
does not have a site except on the boundary.
We also know that $W_d^{(n)}(l_x,l_y)=O(1)$ for $l_x=l_y=2$. 
The cluster of polymers that contributes to its lowest order term 
is composed of $n$ exceptional polymers each of which are composed 
only of a single frozen site.
We can see that $W_d^{(n)}(l_x,l_y)$ for the other pairs of $l_x$ and $l_y$
has the order of $z^{(l_x+l_y-2)}$.
The cluster of polymers that contributes to its lowest order term 
is composed of a polymer 
that has no frozen site and fits into the lattice with the order of
$z^{(l_x+l_y-2)}$ 
and $n$ exceptional polymers each of which are composed 
only of a single frozen site.
So, in order to generate the large-$q$ expansion of the term in 
the free energy density that is proportional to $\mu^n$,
it is enough to take into account the finite-size lattices that 
satisfy $l_x+l_y-2\le N$ in Eq.(\ref{eq:F_o}).

\section{Series}
The algorithms in the previous two sections 
give the large-$q$ expansion of the free energy density 
at the arbitrary inverse temperature $\beta$. 
The energy cumulants can be obtained by taking its derivative 
with respect to $\beta$.
If we are interested in the energy cumulants only 
at the phase transition point, as is the case here, we can set $Y=1+y$ 
and  $\tilde{Y}=1+\tilde{y}$
(with $y=0$ and $\tilde{y}=0 $ at $\beta=\beta_t$)\cite{Guttmann1993}, 
then in order to obtain the $n$th cumulant 
we have only to keep the expansion with respect to $y$ 
to order $y^n$ as
\begin{equation}
F_d^{(n)}=\left.\frac{d^n}{d\beta^n} F_d(\beta) \right|_{\beta=\beta_t}
             =\sum_m a_{d}^{(n,m)} z^m\;,
\end{equation}
using $\frac{d}{d\beta}=(1+y+z)\frac{d}{dy}$ and
\begin{equation}
F_o^{(n)}=\left.\frac{d^n}{d\beta^n} F_o(\beta) \right|_{\beta=\beta_t}
             =\sum_m a_{o}^{(n,m)} z^m\;,
\end{equation}
using 
$\frac{d}{d\beta}=[1+\tilde{y}+(1+\tilde{y})^2z]\frac{d}{d\tilde{y}}$.

Using this procedure we have calculated the series to order $N=23$ 
in $z$ for the $n$th cumulants with $n=0,1,2$ and
to order $N=21$ for the cumulants with $n=3,\cdots,6$.
The coefficients of the series are listed in Table 1 and 2
for the disordered phase and in Table 3 and 4 
for the ordered phase.
We have applied the finite lattice method both in the ordered 
and disordered phases without relying on the 
duality relations (\ref{eq:Duality_cumu}) for the cumulants. 
We have checked that all of $W_d(l_x,l_y)$ and $W_o(l_x,l_y)$ 
with $l_x+l_y-2\le N$ have the correct order 
in $z$ as described in the previous two sections.
The obtained series for the zeroth and first cumulants
(i.e., the free energy and the internal energy) agree
with the expansion of the exactly known expressions.
The coefficients for $F_o^{(n)}$ agree with those
by Bhattacharya {\em et al.} to order 10.
We have also checked that the series satisfy the duality relations 
(\ref{eq:Duality_cumu}) for the cumulants.

\begin{table}[tb]
\caption{
The large-$q$ expansion coefficients $a_{d}^{(n,m)}$
for the $n$th energy cumulants of the free energy
at $\beta=\beta_t$ in the disordered phase ($n=1$--$4$).
         }
\label{tab:coeffada}
\begin{center}
\begin{tabular}{rrrrr}
\hline
$m$      & $a_{d}^{(1,m)}$
         & $a_{d}^{(2,m)}$
         & $a_{d}^{(3,m)}$ 
         & $a_{d}^{(4,m)}$ \\
\hline
$ 0$ &$       $ &$          $ & $             $ & $                $\\
$ 1$ &$      2$ &$         2$ & $            2$ & $               2$\\
$ 2$ &$      4$ &$        14$ & $           58$ & $             242$\\
$ 3$ &$     -2$ &$        26$ & $          338$ & $            3026$\\
$ 4$ &$      2$ &$       118$ & $         2474$ & $           37186$\\
$ 5$ &$     -6$ &$       250$ & $        11490$ & $          291946$\\
$ 6$ &$      4$ &$       894$ & $        55506$ & $         2068914$\\
$ 7$ &$    -16$ &$      1936$ & $       219404$ & $        12144136$\\
$ 8$ &$      6$ &$      6160$ & $       876384$ & $        65992348$\\
$ 9$ &$    -38$ &$     13538$ & $      3127818$ & $       322997186$\\
$10$ &$       $ &$     39774$ & $     11161914$ & $      1494569130$\\
$11$ &$    -76$ &$     88360$ & $     37125056$ & $      6470376688$\\
$12$ &$    -56$ &$    245188$ & $    122847224$ & $     26851465312$\\
$13$ &$    -96$ &$    547468$ & $    387657840$ & $    106378391164$\\
$14$ &$   -348$ &$   1457976$ & $   1214792028$ & $    407805474096$\\
$15$ &$    156$ &$   3264012$ & $   3679570740$ & $   1510368315660$\\
$16$ &$  -1634$ &$   8410284$ & $  11061186460$ & $   5448643138308$\\
$17$ &$   1946$ &$  18868858$ & $  32420680250$ & $  19136730128506$\\
$18$ &$  -6852$ &$  47391870$ & $  94306690802$ & $  65779858359354$\\
$19$ &$  10744$ &$ 106180532$ & $ 269004235012$ & $ 221297704673276$\\
$20$ &$ -27004$ &$ 261607968$ & $ 761823806996$ & $ 731233034945892$\\
$21$ &$  48492$ &$ 586199668$ & $2124082332388$ & $2373735516063508$\\
$22$ &$-102226$ &$1415498882$ & $$              & $$                \\
$23$ &$ 199386$ &$3174300542$ & $$              & $$                \\
\hline
\end{tabular}
\end{center}
\end{table}
\begin{table}[tb]
\caption{
The large-$q$ expnsion coefficients $a_{d}^{(n,m)}$
for the $n$th enrgy cumulants of the free energy 
at $\beta=\beta_t$ in the disordered phase ($n=5,6$).
         }
\label{tab:coeffadb}
\begin{center}
\begin{tabular}{rrr}
\hline
    $m$  & $a_{d}^{(5,m)}$
         & $a_{d}^{(6,m)}$ \\
\hline
$  0$ & $                   $ & $                     $ \\ 
$  1$ & $                  2$ & $                    2$ \\ 
$  2$ & $                994$ & $                 4034$ \\ 
$  3$ & $              24338$ & $               186626$ \\ 
$  4$ & $             488162$ & $              5995138$ \\ 
$  5$ & $            5985042$ & $            110362330$ \\ 
$  6$ & $           60977034$ & $           1583071794$ \\ 
$  7$ & $          503300684$ & $          17852715736$ \\ 
$  8$ & $         3668770296$ & $         171621163420$ \\ 
$  9$ & $        23690762442$ & $        1432076136578$ \\ 
$ 10$ & $       140396232786$ & $       10720828033434$ \\ 
$ 11$ & $       768231161504$ & $       72985931413600$ \\ 
$ 12$ & $      3949989951344$ & $      459566465021728$ \\ 
$ 13$ & $     19184686377744$ & $     2701568006995708$ \\ 
$ 14$ & $     88891085678652$ & $    14974796851686096$ \\ 
$ 15$ & $    394525773133188$ & $    78777820375010652$ \\ 
$ 16$ & $   1687573606453276$ & $   395842580467193364$ \\ 
$ 17$ & $   6978567585546266$ & $  1908650742049266298$ \\ 
$ 18$ & $  28012258546890218$ & $  8869906520648194890$ \\ 
$ 19$ & $ 109405587397512484$ & $ 39863282911219604012$ \\ 
$ 20$ & $ 416940286121010524$ & $173802434948834045508$ \\ 
$ 21$ & $1553287387425965092$ & $737013278292605504308$ \\ 
\hline
\end{tabular}
\end{center}
\end{table}
\begin{table}[tb]
\caption{
The large-$q$ expansion coefficients $a_{o}^{(n,m)}$
for the $n$th energy cumulants of the free energy 
at $\beta=\beta_t$ in the ordered phase ($n=1$--$4$).
         }
\label{tab:coeffaoa}
\begin{center}
\begin{tabular}{rrrrr}
\hline
    $m$  & $a_{o}^{(1,m)}$
         & $a_{o}^{(2,m)}$
         & $a_{o}^{(3,m)}$ 
         & $a_{o}^{(4,m)}$ \\
\hline
$  0$ &$      2$ &$          $ &$              $ &$                $\\
$  1$ &$       $ &$          $ &$              $ &$                $\\
$  2$ &$     -4$ &$        16$ &$           -64$ &$             256$\\
$  3$ &$      2$ &$        34$ &$          -430$ &$            3778$\\
$  4$ &$     -2$ &$       114$ &$         -2654$ &$           41778$\\
$  5$ &$      6$ &$       254$ &$        -12186$ &$          322670$\\
$  6$ &$     -4$ &$       882$ &$        -57018$ &$         2210982$\\
$  7$ &$     16$ &$      1944$ &$       -224732$ &$        12819264$\\
$  8$ &$     -6$ &$      6128$ &$       -888024$ &$        68657204$\\
$  9$ &$     38$ &$     13550$ &$      -3164682$ &$       333583598$\\
$ 10$ &$       $ &$     39698$ &$     -11243178$ &$      1532324246$\\
$ 11$ &$     76$ &$     88360$ &$     -37363472$ &$      6604807168$\\
$ 12$ &$     56$ &$    245036$ &$    -123377384$ &$     27298396784$\\
$ 13$ &$     96$ &$    547356$ &$    -389128512$ &$    107855738700$\\
$ 14$ &$    348$ &$   1457784$ &$   -1218076500$ &$    412466192928$\\
$ 15$ &$   -156$ &$   3263316$ &$   -3688318020$ &$   1524965526804$\\
$ 16$ &$   1634$ &$   8410596$ &$  -11080768444$ &$   5492850472332$\\
$ 17$ &$  -1946$ &$  18865590$ &$  -32471142890$ &$  19269581858838$\\
$ 18$ &$   6852$ &$  47395762$ &$  -94419894146$ &$  66169209300214$\\
$ 19$ &$ -10744$ &$ 106166828$ &$ -269288597908$ &$ 222430064188868$\\
$ 20$ &$  27004$ &$ 261629456$ &$ -762460849076$ &$ 734462791941548$\\
$ 21$ &$ -48492$ &$ 586145660$ &$-2125652044660$ &$2382881224028156$\\
$ 22$ &$ 102228$ &$1415594740$ &$$               &$$                \\
$ 23$ &$-199212$ &$3174081000$ &$$               &$$                \\
\hline
\end{tabular}
\end{center}
\end{table}
\begin{table}[tb]
\caption{
The large-$q$ expansion coefficients $a_{n,m}^{(o)}$
for the $n$th energy cumulants of the free energy
at $\beta=\beta_t$ in the ordered phase ($n=5,6$).
         }
\label{tab:coeffaob}
\begin{center}
\begin{tabular}{rrr}
\hline
$m$   & $           a_{o}^{(5,m)}$ & $            a_{o}^{(6,m)}$ \\
\hline
$  0$ & $                    $ & $                     $ \\ 
$  1$ & $                    $ & $                     $ \\ 
$  2$ & $               -1024$ & $                 4096$ \\ 
$  3$ & $              -29518$ & $               218914$ \\ 
$  4$ & $             -556382$ & $              6813714$ \\ 
$  5$ & $            -6773994$ & $            126069374$ \\ 
$  6$ & $           -67122114$ & $           1774583142$ \\ 
$  7$ & $          -546094604$ & $          19774354944$ \\ 
$  8$ & $         -3918393456$ & $         187361651588$ \\ 
$  9$ & $        -25037212842$ & $        1545876302510$ \\ 
$ 10$ & $       -146961943266$ & $       11451708807878$ \\ 
$ 11$ & $       -798499755344$ & $       77296110810160$ \\ 
$ 12$ & $      -4080741048224$ & $      483066658347536$ \\ 
$ 13$ & $     -19726182681504$ & $     2822025943834956$ \\ 
$ 14$ & $     -91033421848212$ & $    15558449192797824$ \\ 
$ 15$ & $    -402728474968788$ & $    81476650613568516$ \\ 
$ 16$ & $   -1717926911825116$ & $   407801271671111676$ \\ 
$ 17$ & $   -7087982430811466$ & $  1959732852092226390$ \\ 
$ 18$ & $  -28396321463796218$ & $  9080903533904186662$ \\ 
$ 19$ & $ -110725077538793524$ & $ 40709407799692233428$ \\ 
$ 20$ & $ -421377562607301884$ & $177104383529723970956$ \\ 
$ 21$ & $-1567944342623768692$ & $749588012269824549980$ \\ 
\hline
\end{tabular}
\end{center}
\end{table}

The $n$th magnetization cumulant of the free energy at $\beta=\beta_t$ and 
$h=0$ is defined by
\begin{equation}
  M_{d,o}^{(n)}= \left. (\frac{q}{q-1})^n
  \frac{\partial^n}{\partial h^n}F_{d,o}\right|_{h=0,\beta=\beta_t} 
  -(\frac{1}{q-1})\delta_{n,1} 
  =\sum_m b_{d,o}^{(n,m)} z^m 
  \label{eq:cumulantm}
\end{equation}
It can be calculated from the expansion series of the free energy in $\mu$ 
to order $\mu^n$
using $\frac{d}{dh}=(1+\mu)\frac{d}{d\mu}$.
We have calculated the series for the $n$th cumulants with $n=1,2,3$
to order $N=21$ in $z$. 
The coefficients of the series are listed in Table 5 and 6
for the disordered and ordered phases, respectively.
The series for the first cumulant
(i.e., the magnetization) in the disordered phase is vanishing 
and not listed here.
The series for the first cumulant
in the ordered phase agrees 
with the expansion of its exactly known expression.
\begin{table}[t]
\caption{
The large-$q$ expansion coefficients $b_d^{(n,m)}$
for the $n$th magnetization cumulants of the free energy
at $\beta=\beta_t$ in the disordered phase ($n=2$--$3$)
         }
\label{tab:coeffbd}
\begin{center}
\begin{tabular}{rrr}
\hline
    $m$  &  $b_d^{(2,m)}$
         &  $b_d^{(3,m)}$ \\
\hline
$  0$ & $            $ & $               $ \\ 
$  1$ & $            $ & $               $ \\ 
$  2$ & $           1$ & $              1$ \\ 
$  3$ & $           4$ & $             12$ \\ 
$  4$ & $          21$ & $            120$ \\ 
$  5$ & $          72$ & $            804$ \\ 
$  6$ & $         289$ & $           4943$ \\ 
$  7$ & $         920$ & $          25764$ \\ 
$  8$ & $        3197$ & $         126310$ \\ 
$  9$ & $        9632$ & $         564780$ \\ 
$ 10$ & $       30589$ & $        2412549$ \\ 
$ 11$ & $       88412$ & $        9714144$ \\ 
$ 12$ & $      264037$ & $       37758464$ \\ 
$ 13$ & $      738508$ & $      140800740$ \\ 
$ 14$ & $     2109461$ & $      510552463$ \\ 
$ 15$ & $     5744540$ & $     1794840960$ \\ 
$ 16$ & $    15861797$ & $     6167896002$ \\ 
$ 17$ & $    42249692$ & $    20689927908$ \\ 
$ 18$ & $   113575441$ & $    68107803125$ \\ 
$ 19$ & $   296922168$ & $   219877596348$ \\ 
$ 20$ & $   781015901$ & $   698663386396$ \\ 
$ 21$ & $  2009401452$ & $  2184569257680$ \\ 
\hline
\end{tabular}
\end{center}
\end{table}
\clearpage
\begin{table}[t]
\caption{
The large-$q$ expansion coefficients $b_o^{(n,m)}$
for the $n$th magnetization cumulants of the free energy
at $\beta=\beta_t$ in the ordered phase ($n=1$--$3$)
         }
\label{tab:coeffbo}
\begin{center}
\begin{tabular}{rrrr}
\hline
    $n$  & $b_o^{(1,m)}$
         & $b_o^{(2,m)}$
         & $b_o^{(3,m)}$ \\
\hline
$  0$ & $      1$ & $           $ & $               $ \\ 
$  1$ & $       $ & $           $ & $               $ \\ 
$  2$ & $     -1$ & $          1$ & $             -1$ \\ 
$  3$ & $       $ & $          4$ & $            -12$ \\ 
$  4$ & $     -3$ & $         23$ & $           -122$ \\ 
$  5$ & $       $ & $         80$ & $           -828$ \\ 
$  6$ & $     -9$ & $        327$ & $          -5159$ \\ 
$  7$ & $       $ & $       1048$ & $         -27156$ \\ 
$  8$ & $    -27$ & $       3649$ & $        -134398$ \\ 
$  9$ & $       $ & $      11020$ & $        -605496$ \\ 
$ 10$ & $    -82$ & $      34902$ & $       -2604438$ \\ 
$ 11$ & $       $ & $     100840$ & $      -10546968$ \\ 
$ 12$ & $   -254$ & $     299901$ & $      -41207005$ \\ 
$ 13$ & $       $ & $     837240$ & $     -154329576$ \\ 
$ 14$ & $   -804$ & $    2380574$ & $     -561777935$ \\ 
$ 15$ & $       $ & $    6465720$ & $    -1981489368$ \\ 
$ 16$ & $  -2596$ & $   17772083$ & $    -6829373776$ \\ 
$ 17$ & $       $ & $   47195596$ & $   -22967338956$ \\ 
$ 18$ & $  -8520$ & $  126320724$ & $   -75775684452$ \\ 
$ 19$ & $       $ & $  329203956$ & $  -245115502776$ \\ 
$ 20$ & $ -28311$ & $  862396671$ & $  -780219055115$ \\ 
$ 21$ & $       $ & $ 2211793232$ & $ -2443319594304$ \\ 
\hline
\end{tabular}
\end{center}
\end{table}
\clearpage

\section{Analysis}
In the first part of this section 
we analyze the large-$q$ series for the energy cumulants 
presented in the previous section.
The series of the magnetization cumulants are analyzed in the rest part
of this section.
The asymptotic behavior of the energy cumulants at $\beta=\beta_t$ 
for $q \to 4_+$ was conjectured 
by Bhattacharya {\it et al.}\cite{Bhattacharya1994} as follows. 
It is well known that there is an asymptotic  
relation between the energy cumulants and the correlation length $\xi$ 
for $\beta\rightarrow\beta_t$ at $q=4$ as
\begin{equation}
 F_d^{(n)},(-1)^n F_o^{(n)} 
 \sim A \frac{\Gamma\left( n-\frac{4}{3}\right)}
 {\Gamma \left(\frac{2}{3}\right)} \xi^{3n/2-2}, \label{eq:relation}
\end{equation}
with the constant $A$ that is independent of $n$. 
This relation comes from the fact that
the critical exponents of the correlation length and the specific heat 
in the second order phase transition
are $\nu=2/3$ and $\alpha=2/3$, respectively, for $q=4$.
Their conjecture claims that the relation (\ref{eq:relation})
is also held as the asymptotic relation between the energy cumulants 
and the correlation length at the first order phase transition point
for $q\rightarrow 4_+$.
Then, from the known asymptotic behavior of the correlation length 
at the first order phase transition point 
for $q\to 4_+$ as\cite{Borgs}
\begin{equation}
\xi \sim \frac{1}{8\sqrt{2}}x\;, \label{eq:correlation}
\end{equation}
where $x\equiv \exp{\left( \frac{\pi^2}{2\theta}\right)}$
with $\theta$ defined by $2\cosh{\theta}\equiv \sqrt{q}$
($\theta \sim \sqrt{q-4}/2$ for $q\to 4_+$), 
the conjecture implies that the asymptotic form 
of the energy cumulants in the limit of $q\rightarrow 4_+$ is given by
\begin{equation}
  F_d^{(n)}, (-1)^n F_o^{(n)} \sim \alpha B^{n-2}
  \frac{ \Gamma \left( n - \frac{4}{3} \right) }{ \Gamma
    \left( \frac{2}{3} \right) } x^{3n/2-2}\;. \label{eq:asymp_form}
\end{equation}
We notice that the constants $\alpha$ and $B$ should be 
independent of $n$
and from the duality relations (\ref{eq:Duality_cumu})
that they should be common to the ordered and disordered phases.

To analyze the large-$q$ series for these quantities, 
Bhattacharya {\it et al.} used a sophisticated
method.  
Let $f(z)$ be a function that has the asymptotic behavior 
of the right hand side of Eq.(\ref{eq:asymp_form}).  
Then $\theta \ln{f(z)}$ is a regular function of $z$ 
except for the possible terms proportional 
to the inverse powers of $x$, which vanishes rapidly when $\theta$ 
approaches zero.
Although $\theta$ itself cannot be expanded as the power
series of $z$ since $\theta$ diverge at $z=0$, 
the quantity $1 - e^{-\theta} = 1 - 2z / (\sqrt{1 - 4z^2} + 1)$ 
behaves like $\theta$ for $\theta\to 0$ 
and can be expanded as a power series of $z$.
Then one may expect that $(1-e^{-\theta})\ln{f(z)}$ will be 
a smooth function of $z$ and its Pad\'e approximant will give
a convergent result for $q\to 4_+$.
In fact using the 10th order series Bhattacharya {\it et al.} 
found that the Pad\'e approximants
of $F^{(2)}$ give convergent results for $q \gtsim 7$. 
The obtained results were more precise than the results of the Monte 
Carlo simulations\cite{Janke} and the low-(or high-)temperature 
series\cite{Bhanot1993,Briggs,Arisue1997} as mentioned 
in the introduction
and exhibits rather flat behavior of $F^{(2)}/x$ 
around $7\ltsim q\ltsim 10$, 
suggesting the correctness of the asymptotic behavior 
(\ref{eq:asymp_form}) for $n=2$ with $\alpha\sim 0.76$.
If we use our longer 23rd order series of $F^{(2)}$
to repeat these Pad\'e analyses, 
we find that the convergent region of $q$ 
is extended down to $q\sim 5$, with the accuracy of a few percent 
at $q=5$ and the data of $F^{(2)}/x$ is rather flat 
for $5\ltsim q\ltsim 10$, 
which gives a stronger support to the conjecture that $F^{(2)}$ 
is approaching its asymptotic form (\ref{eq:asymp_form}).

We, however, adopt here another method of the analysis that
is simpler and more efficient. 
The latent heat ${\cal L}$ has the exact asymptotic form 
as\cite{Bhattacharya1994} 
$$
{\cal L} \sim 3\pi x^{-1/2}\;, \label{eq:latentheat}
$$
so, if $F^{(n)}$ has the asymptotic form of Eq.(\ref{eq:asymp_form}), 
the product $F^{(n)}{\cal L}^{p}$ is a smooth function of $z$
for $q\to 4_+$ when $p=3n-4$, 
while it behaves like some positive or negative powers of $x$ and 
has an essential singularity 
with respect to $z$ at $q=4$ when $p\ne 3n-4$.
Thus if we try the Pad\'e approximation of $F^{(n)}{\cal L}^{p}$
it will lead to a convergent result for $p=3n-4$, 
while the approximants will scatter when $p$ is off this value.
In the Pad\'e approximation of $F^{(n)}{\cal L}^{p}$ 
the series for the ordered and disordered phases should be treated 
in a slightly different way, 
corresponding to the difference of the lowest order of the series; 
the series for the ordered phase starts from the order of $z^2$, 
while the series for the disordered phase starts 
from the order of $z$.  
Thus we should solve the following equation:
\begin{eqnarray}
   F^{(n)} {\cal L}^{3n-4} = z^r P_M(z)/Q_L(z) + O(z^{M+L+r+1})\;,
\end{eqnarray}
where $r$ is 2 for the ordered phase and 1 for the disordered phase, 
and $P_M(z)$ and $Q_L(z)$ are polynomials of order $M$ and 
$L$ respectively with $M + L + r=N$ 
where $N$ is the truncated order of the series.
Fig.11 is the result of this analysis of $F_d^{(2)}{\cal L}^{p}$ 
for $q=4$.
We see very clearly that the Pad\'e approximants have the best 
convergence at $p=2.00(1)$ and the convergence becomes bad rapidly 
when $p$ leaves this value.
All the other approximants for $F_{d,o}^{(n)}{\cal L}^{p}$ 
have similar behaviors with respect to $p$ with the best convergence at
$p=3n-4$, although we do not show them in figures.
These results are enough to convince us 
that $F_{d,o}^{(n)}$ has the asymptotic behavior 
that is proportional to $x^{3n/2-2}$.
\begin{figure}[tb]
\epsfxsize=7.4cm
\epsffile{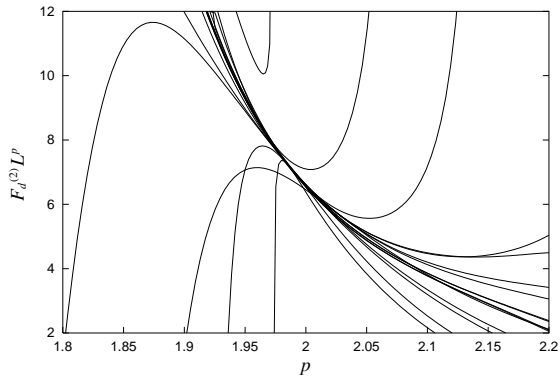}
\caption{
Pad\'e approximants of $F_d^{(2)}{\cal L}^p$ at $q=4$ with $M\ge 9$ 
and $L\ge 9$ plotted versus $p$. 
         }
\label{fig11}
\end{figure}

Then we calculate the amplitude of $F^{(n)}$ for each $q$
by taking the Pad\'e approximants of $F^{(n)}{\cal L}^{3n-4}$ 
and multiplying ${\cal L}^{4-3n}$ that is calculated 
from its exact expression.
Usually a diagonal approximant (that is, $M=L$) or its neighborhood 
produces good estimates. 
We, however, adopt here all the approximants 
with $M \ge 8$ and $L \ge 8$, 
since we have found in this case that each of these approximants 
is scattered almost randomly around their average value 
and we have found no reason to give a special position 
only to the diagonal approximant.
Some denominators of the approximants have zero at $z<1/2$ ($q>4$) 
and we have excluded such approximants.
The results for $n=2,\cdots,6$ in the ordered and disordered phases 
are listed in Table 7--10 for integer values of $q\ge 5$, 
with the second energy cumulant transformed to the specific heat
by multiplying $\beta_t^2$. 
In Figs. 12-16, we also plot $|F^{(n)}| / x^{\frac{3n}{2}-2}$ 
versus $\theta$ for $n=$2--6, respectively.
In these figures the dotted and dashed lines indicate 
the errors for the disordered and ordered phases, respectively.
In these tables and figures the given errors represent 
the magnitude of the fluctuation 
in the data of the Pad\'e approximants.

In any case, the estimates have surprisingly high accuracy. 
Our data are orders of magnitude more precise than 
the result of the Monte Carlo simulation 
by Janke and Kappler\cite{Janke} for $q=10,15,20$ 
(also listed in Table 7--10),
which was performed carefully in large enough lattices checking 
the finite size effects, 
and they are consistent with each other 
except for small differences in the data of the 5th and 6th cumulants 
for the ordered phase.
On the other hand, the results 
of the low-temperature series\cite{Arisue1997} 
(also listed in Table 7 and 8) are inconsistent with these.
This would be because we do not have the correct knowledge 
(or conjecture) of the possible singularity structure of the cumulants 
at some $\beta\le\beta_t$ in extrapolating the low-temperature series 
toward $\beta=\beta_t$.
We should emphasize that the value of the
specific heat has an accuracy of about 0.1 percent at $q=5$ 
where the correlation length is as large as 2500 lattice 
spacing\cite{Borgs} and there is no previous data to be compared with, 
and of a few percent even at the limit of $q=4$ 
where the correlation length diverges.
The accuracy is better by a factor of more than 10 
when compared with the result of using the Pad\'e approximation for 
$(1-e^{-\theta})\ln{F^{(n)}}$. 
The reason for this is the following.
Let us suppose that the Pad\'e approximants give estimates of
$F^{(n)}{\cal L}^{3n-4}=c(1\pm \delta)$ and 
$(1-e^{-\theta})\ln{F^{(n)}}=c^{\prime}(1\pm \delta^{\prime})$, 
then the estimate of $F^{(n)}$ from the former would also have 
an ambiguity factor of $1\pm \delta$, 
while the estimate of $F^{(n)}$ from the latter would have 
an ambiguity factor of $e^{\pm c^{\prime}\delta^{\prime}/\theta}$, 
which grows up for the plus sign and shrinks to zero 
for the minus sign when $\theta\to 0(q\to 4_+)$. 

Next we determine the constant $\alpha$ and $\beta$ 
in the asymptotic form (\ref{eq:asymp_form}) for $q\to 4_+$.
The constants $\alpha$ can be determined by the value of $F^{(2)}/x$ 
in the limit of $q\to 4_+$. 
We read from Fig.12 that
$$ \alpha = 0.073 \pm 0.002\;. $$
On the other hand, the constant $B$ is given
by the asymptotic value of the following combination
for $n\ge 3$:
\begin{equation}  
 \left\{
   \frac{ \Gamma \left( n - \frac{4}{3} \right) | F^{(n)} | }{
    \Gamma \left( \frac{2}{3} \right) F^{(2)}}
  \right\} ^{\frac{1}{n-2}} x^{-\frac{3}{2}} \;. 
   \label{eq:constant_B}
\end{equation}  
Their asymptotic values should be common to each $n$.
The behavior of the quantity in Eq.(\ref{eq:constant_B})
for the ordered and disordered phases are given in Fig.17 and 18 
respectively.
The graphs for $n=3$ and $n=4$ 
show clear convergence even at $\theta =0$ giving the value of $B$ 
very close to each other, that is, $B\sim 0.38(3)$
We also notice that, 
although the results for $n=5$ and $n=6$ have considerably large 
errors 
in the small $\theta$ region so we cannot say anything definite, 
the values of the quantity in Eq.(\ref{eq:constant_B}) at the points 
where the errors begin to grow are also close 
to this value of 0.38(3). 
Thus we may estimate the constant $B$ as
$$ B = 0.38 \pm 0.03\;. $$
In conclusion the Pad\'e data of the large-$q$ series for the energy cumulants
convince us that 
the conjecture (\ref{eq:asymp_form}) for the asymptotic form 
is true at least for $n=2,3$ and $4$ 
and suggest its correctness for $n=5$ and $6$.
\begin{table}[tb]
\caption{Estimates of the specific heat $C_d$ for the disordered phase
  in comparison with the low-temperature series and the Monte Carlo
  simulations.}
\label{tab:tab5}
\begin{center}
\begin{tabular}{llll}
\hline
    $q$  & \multicolumn{1}{c}{large-$q$ }
         & \multicolumn{1}{c}{low-temp. }
     & \multicolumn{1}{c}{Monte Carlo}\\
\hline
 5 & 2889.1(41)          &            & \\
 6 & 205.930(55)         &            & \\
 7 & 68.7370(52)         & 68.01(47)  & \\
 8 & 36.9335(11)         & 30.17(6)   & \\
 9 & 24.58756(35)        & 20.87(5)   & \\
10 & 18.38542(17)        & 16.294(34) & 18.437(40) \\
15 & 8.6540342(98)       & 8.388(4)   & 8.6507(57) \\
20 & 6.132159678(47)     & 6.0645(69) & 6.1326(4) \\
30 & 4.29899341467(73)   & 4.2919(51) & \\
\hline
\end{tabular}
\end{center}
\end{table}
\begin{table}
\caption{Estimates of the specific heat $C_o$ for the ordered phase
  in comparison with the low-temperature series and the Monte Carlo
  simulations.}
\label{tab:tab6}
\begin{center}
\begin{tabular}{llll}
\hline
    q  & \multicolumn{1}{c}{large-$q$ }
       & \multicolumn{1}{c}{low-temp.}
       & \multicolumn{1}{c}{Monte Carlo}\\
\hline
 5 & 2885.8(34)          &            & \\
 6 & 205.780(32)         &            & \\
 7 & 68.5128(22)         & 52.98(61)  & \\
 8 & 36.62347(31)        & 28.47(47)  & \\
 9 & 24.203436(65)       & 20.03(27)  & \\
10 & 17.937801(18)       & 15.579(83) & 17.989(40) \\
15 & 7.99645871(22)  & 7.721(28)  & 7.999(3) \\
20 & 5.360768767(13)     & 5.2913(53) & 5.3612(4) \\
30 & 3.41289525542(28)   & 3.4012(10) & \\
\hline
\end{tabular}
\end{center}
\end{table}
\begin{table}
\caption{Estimates of the energy cumulants $F_d^{(n)}$
  for the disordered phase ($n=3$--$6$). 
  The values in the brackets are the data of the Monte Carlo
  simulations.}
\label{tab:tab7}
\begin{center}
\begin{tabular}{lllll}
\hline
    q  & \multicolumn{1}{c}{$F_d^{(3)}$}
       & \multicolumn{1}{c}{$F_d^{(4)}$}
       & \multicolumn{1}{c}{$F_d^{(5)}$}
       & \multicolumn{1}{c}{$F_d^{(6)}$}\\
\hline
 5 & 2.1597(27) $\times 10^9$  & 7.4380(99) $\times 10^{15}$  
   & 4.90(39) $\times 10^{22}$    & 4.9(15) $\times 10^{29}$ \\
 6 & 2.04752(87) $\times 10^6$ & 1.07367(47) $\times 10^{11}$ 
   & 1.092(30) $\times 10^{16}$   & 1.68(38) $\times 10^{21}$ \\
 7 & 9.8404(16) $\times 10^4$  & 8.2877(14) $\times 10^8$     
   & 1.363(14) $\times 10^{13}$   & 3.44(52) $\times 10^{17}$ \\
 8 & 16420.4(11)               & 4.69139(37) $\times 10^7$  
   & 2.625(11) $\times 10^{11}$   & 2.30(20) $\times 10^{15}$ \\
 9 & 4880.80(16)               & 6.69421(26) $\times 10^6$  
   & 1.8024(36) $\times 10^{10}$  & 7.67(37) $\times 10^{13}$ \\
10 & 2002.017(32)              & 1.600281(33) $\times 10^6$ 
   & 2.5155(25) $\times 10^9$   & 6.29(16) $\times 10^{12}$ \\
   & [2015(26)]                & [1.583(64) $\times 10^6$]  
   & [2.40(26) $\times 10^9$]   & [5.8(18) $\times 10^{12}$] \\
15 & 176.78507(17)             & 32323.501(59)     
   & 1.168612(78) $\times 10^7$ & 6.779(13) $\times 10^9$ \\
   & [176.01(76)]              & [2.93(20) $\times 10^4$] 
   & [1.170(59) $\times 10^7$]  & [7.2(11) $\times 10^9$] \\
20 & 54.8121419(72)            & 4904.2443(16)     
   & 870078.5(88)   & 2.47889(69) $\times 10^8$ \\
   & [54.7(4)]                 & [4905(89)]        
   & [8.42(32) $\times 10^5$] & [2.33(16) $\times 10^8$]  \\
30 & 15.88750157(25)           & 667.244592(17)    
   & 55619.490(41)  & 7.44878(13) $\times 10^6$ \\
\hline
\end{tabular}
\end{center}
\end{table}
\begin{table}
\caption{
Absolute value of the estimates of the energy cumulants $F_o^{(n)}$ 
for the ordered phase ($n=3$--$6$). 
The values in the brackets are the data of the Monte Carlo simulations.
}
\label{tab:tab8}
\begin{center}
\begin{tabular}{lllll}
\hline
    q  & \multicolumn{1}{c}{$F_o^{(3)}$}
       & \multicolumn{1}{c}{$F_o^{(4)}$}
       & \multicolumn{1}{c}{$F_o^{(5)}$}
       & \multicolumn{1}{c}{$F_o^{(6)}$}\\
\hline
 5 & 2.1577(84) $\times 10^9$ & 7.368(33) $\times 10^{15}$  
   & 5.07(21)     $\times 10^{22}$ & 5.15(46)  $\times 10^{29}$ \\
 6 & 2.0467(28) $\times 10^6$ & 1.0698(22) $\times 10^{11}$ 
   & 1.105(13)    $\times 10^{16}$ & 1.775(60) $\times 10^{21}$ \\
 7 & 9.8469(57) $\times 10^4$ & 8.2777(91) $\times 10^8$  
   & 1.3691(65)   $\times 10^{13}$ & 3.586(52) $\times 10^{17}$ \\
 8 & 16460.8(46)      & 4.6939(32) $\times 10^7$ 
   & 2.6335(57)   $\times 10^{11}$ & 2.356(16) $\times 10^{15}$ \\
 9 & 4905.23(73)      & 6.7098(35) $\times 10^6$ 
   & 1.8084(19)   $\times 10^{10}$ & 7.799(27) $\times 10^{13}$ \\
10 & 2018.31(17)      & 1.6071(11) $\times 10^6$ 
   & 2.5268(14)   $\times 10^9$ & 6.367(12)  $\times 10^{12}$ \\
   & [2031(26)]       & [1.55(22) $\times 10^6$] 
   & [1.98(51) $\times 10^9$]   & [2.8(14) $\times 10^{12}$] \\
15 & 181.4445(19)     & 32878.80(86)     
   & 1.185486(57) $\times 10^7$ & 6.8744(12) $\times 10^9$ \\
   & [180.67(76)]     & [2.93(20) $\times 10^4$] 
   & [0.92(15) $\times 10^7$]   & [4.5(13) $\times 10^9$]  \\
20 & 57.05595(23)     & 5054.1460(86)    
   & 892352.53(68)   & 2.538205(72) $\times 10^8$ \\
   & [57.09(29)]      & [4.79(22) $\times 10^3$] 
   & [7.37(70) $\times 10^5$] & [1.38(26) $\times 10^8$] \\
30 & 16.7446548(22)   & 702.788233(56)   
   & 58121.979(30)   & 7.760343(17) $\times 10^6$ \\
\hline
\end{tabular}
\end{center}
\end{table}
\begin{figure}[tb]
\epsfxsize=7.4cm
\epsffile{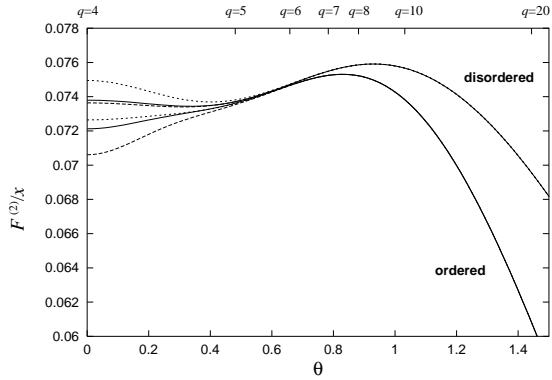}
\caption{
The behavior of the ratio of the second energy cumulants 
$F_{d,o}^{(2)}$ to $x$.
         }
\label{fig12}
\end{figure}
\begin{figure}[tb]
\epsfxsize=7.4cm
\epsffile{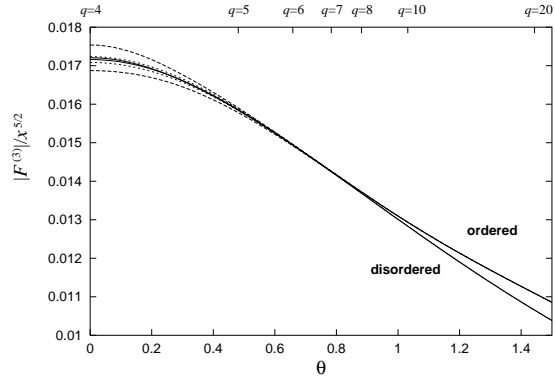}
\caption{
The behavior of the ratio of the absolute value of
the third energy cumulants 
$|F_{d,o}^{(3)}|$ to $x^{5/2}$.
         }
\label{fig13}
\end{figure}
\begin{figure}[tb]
\epsfxsize=7.4cm
\epsffile{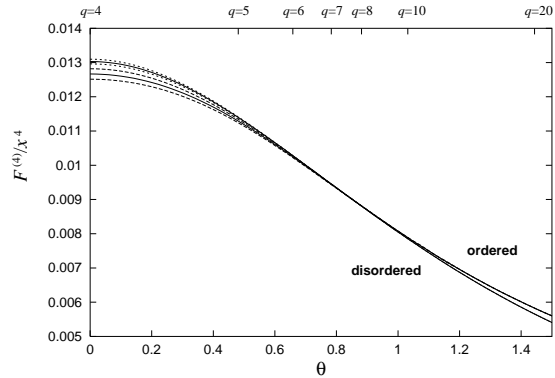}
\caption{
The behavior of the ratio of
the fourth energy cumulants 
$F_{d,o}^{(4)}$ to $x^4$.
         }
\label{fig14}
\end{figure}
\begin{figure}[tb]
\epsfxsize=7.4cm
\epsffile{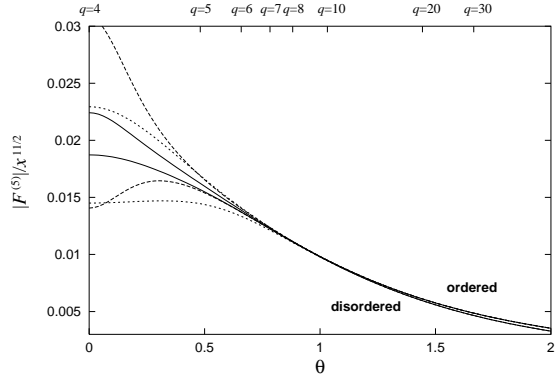}
\caption{
The behavior of the ratio of the absolute value of
the fifth energy cumulants 
$|F_{d,o}^{(5)}|$ to $x^{11/2}$.
         }
\label{fig15}
\end{figure}
\begin{figure}[tb]
\epsfxsize=7.4cm
\epsffile{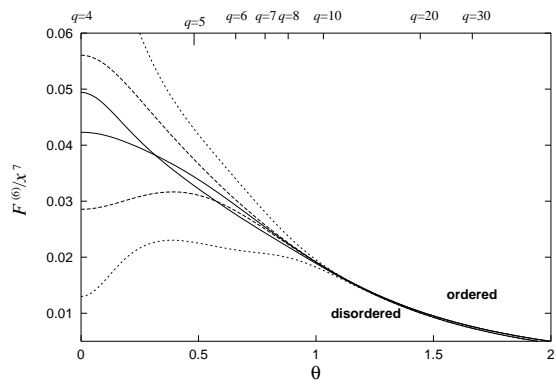}
\caption{
The behavior of the ratio of
the sixth energy cumulants 
$F_{d,o}^{(6)}$ to $x^7$.
         }
\label{fig16}
\end{figure}
\clearpage

\begin{figure}[tb]
\epsfxsize=7.4cm
\epsffile{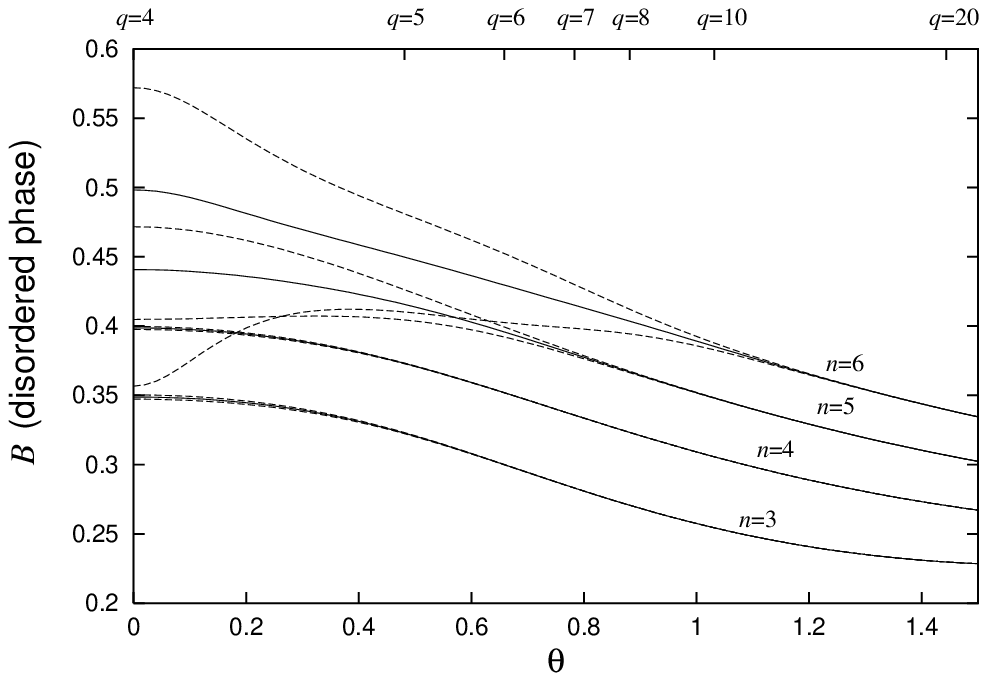}
\caption{
The behavior of the comibined quantity of 
Eq.(21) in the disordered phase.
}
\label{fig17}
\end{figure}
\begin{figure}[tb]
\epsfxsize=7.4cm
\epsffile{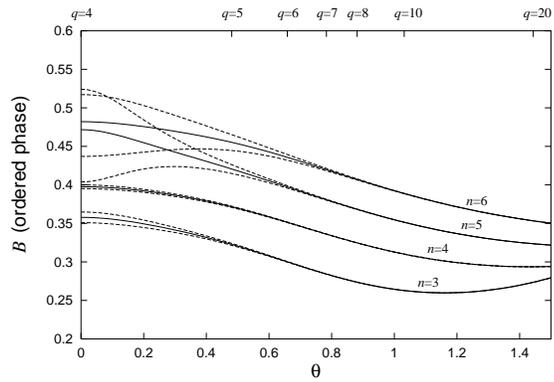}
\caption{
The behavior of the comibined quantity of 
Eq.(21) in the ordered phase.
         }
\label{fig18}
\end{figure}
\clearpage

Now, we analyze the series of the magnetization cumulants.
The $n$th magnetization cumulants $M^{(n)}$ 
is well known to behave for $\beta\to\beta_t$ at $q=4$ as
\begin{equation}
M_{d}^{(n)},(-1)^nM_{o}^{(n)} \sim A_{d,o}^{(n)} (\xi)^{\frac{15}{8}n-2}
\end{equation}
and we can make a conjecture parallel to the conjecture 
for the energy cumulants by Bhattacharya {\it et al.}: 
this relation will also be kept in the limit $q\to 4_+$ with $\beta=\beta_t$, 
which implies using the asymptotic behavior of the correlation length 
in Eq.(\ref{eq:correlation}) in this limit that
\begin{equation}
M_{d}^{(n)},(-1)^nM_{o}^{(n)} \sim B_{d,o}^{(n)} x^{\frac{15}{8}n-2}\;.
         \label{eq:asymp_m}
\end{equation}
Therefore we use the same method of analysis as in the case of 
the energy cumulants.

At first, we have checked that the approximants for $M_{d,o}^{(n)}{\cal L}^p$
have the best convergent results just at $p=(15n-16)/4$, and next
we have estimated the magnetization cumulants
from the Pad\'e approximants for $M_{d,o}^{(n)}{\cal L}^{(15n-16)/4}$
whose $M$ and $N$ are greater than 8 and the denominators does not
possess zero at $0<z<1/2$.
The results for $n=2$ and $3$ in the ordered and disordered phases 
are listed in Table 11 for integer values of $q\ge 5$.
The estimates of the magnetization cumulants have also surprisingly high 
accuracy. 
Our data are orders of magnitude more precise than 
the result of the Monte Carlo simulation 
by Janke and Kappler\cite{Janke} for $q=10,15,20$ (also listed in Table 11).
In Figs. 19 and 20, we plot $|M^{(n)}| / x^{(15n-16)/8}$ versus
$\theta$ for $n=2$ and 3, respectively.  
The constants $B_{d,o}^{(2)}$
and $B_{d,o}^{(3)}$ in the asymptotic behavior in Eq.(\ref{eq:asymp_m})
might be different from the values at $\theta=0$ in the figures. 
The reason is the following. 
The asymptotic form (\ref{eq:latentheat}) of the latent heat ${\cal L}$ 
for $q\to 4_+$ is multiplied by the correction factor that is 
an even function of $\theta$\cite{Baxter1973} while 
the asymptotic form (\ref{eq:latentheat}) of the magnetization
cumulants $M^{(n)}$ is plausible to be multiplied 
by the correction factor that is not an even function of $\theta$, 
since in the case of the first cumulant(i.e., the magnetization) 
the correction factor is not an even function of $\theta$\cite{Baxter1973}. 
If it would be the case, the combination $|M^{(n)}| / x^{(15n-16)/8}$ would 
not be an even function of $\theta$, then it could not be approximated well 
arround $\theta=0$ by the function of $z$, 
which is an even function of $\theta$.
(On the contrary in the case of the energy cumulants 
the asymptotic form (\ref{eq:asymp_form}) is plausible to be multiplied 
by the correction factor that is an even function of $\theta$,
since the latent heat is just the difference 
between the two first energy cumulants. So the combination 
$F^{(n)}{\cal L}^{3n-4}$ is expected to be well approximated 
by the Pad\'e approximants in $z$ even around $\theta=0$.)
Therefore, we will give the prediction of the values of $B_{d,o}^{(n)}$
based on the assumption
that $M^{(n)} / x^{(15n-16)/8}$ varies as a linear function
of $\theta$ around $\theta=0$ and one can obtain the values at
$\theta=0$ by linear extrapolation from the curves 
around $\theta \sim 0.5$ in Fig.19 and 20;
\begin{eqnarray*}
  & B_d^{(2)} = (2.05 \pm 0.05) \times 10^{-3}, \quad 
  & B_o^{(2)} = (1.72 \pm 0.02) \times 10^{-3}  \\
  & B_d^{(3)} = (8.0  \pm 0.1)  \times 10^{-5},  \quad 
  & B_o^{(3)} = (8.57  \pm 0.05)  \times 10^{-5}.
\end{eqnarray*}

\begin{table}[tb]
\caption{Absolute value of the estimates of the magnetic
cumulants $M_{d,o}^{(2)}$ and $M_{d,o}^{(3)}$.
The values in the brackets are the data of the Monte Carlo simulations.}
\label{tab:tab9}
\begin{center}
\begin{tabular}{lll}
\hline
q  & $M_d^{(2)}$        & $M_o^{(2)}$        \\
\hline
 5 & 9.128(31) $\times 10^4$ & 9.011(23) $\times 10^{4}$  \\
 6 & 628.21(37)         & 666.46(39)        \\
 7 & 70.5446(94)        & 77.304(13)        \\
 8 & 19.35941(71)       & 21.5251(13)       \\
 9 & 8.057894(95)       & 9.01050(24)       \\
10 & 4.232764(18)       & 4.738206(66)      \\
   &                    & [4.744(42)]       \\
15 & 0.730421382(66)    & 0.80569715(69)    \\
   &                    & [0.8052(28)]      \\
20 & 0.3093656816(19)   & 0.335564221(11)   \\
   &                    & [0.33551(81)]     \\
30 & 0.121320123911(18) & 0.12867774409(11) \\
\hline
\end{tabular}
\end{center}
\begin{center}
\begin{tabular}{lll}
\hline
 q & $M_d^{(3)}$      & $-M_o^{(3)}$            \\
\hline
 5 & 8.445(28)  $\times 10^{11}$ & 9.336(21)  $\times 10^{11}$ \\
 6 & 3.3527(36) $\times 10^{7}$  & 3.7654(41) $\times 10^{7}$  \\
 7 & 4.0328(17) $\times 10^{5}$  & 4.5420(35) $\times 10^{5}$  \\
 8 & 2.6335(57) $\times 10^{4}$  & 3.3477(59) $\times 10^{4}$  \\
 9 & 5116.70(50)     & 5.7293(28) $\times 10^{3}$ \\
10 & 1405.906(76)    & 1566.25(99)    \\
   &                 & [1532(74)]     \\
15 & 42.43916(31)    & 46.25783(43)   \\
   &                 & [45.3(1.0)]    \\
20 & 7.94284(13)     & 8.523886(23)   \\
   &                 & [8.44(33)]     \\
30 & 1.364338789(50) & 1.43676408(17) \\
\hline
\end{tabular}
\end{center}
\end{table}
\clearpage
\begin{figure}[tb]
\epsfxsize=7.4cm
\epsffile{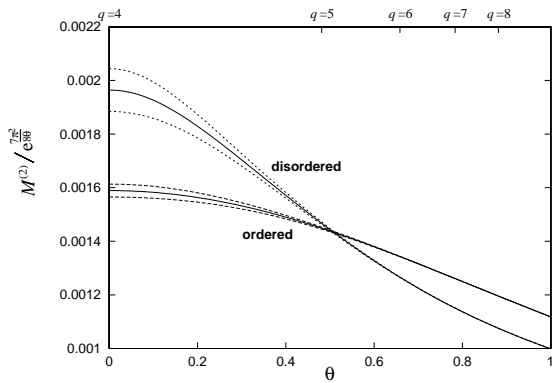}
\caption{
The behavior of the ratio 
the second magnetization cumulants 
$M_{d,o}^{(2)}$ to $x^{7/4}$.
         }
\label{fig19}
\end{figure}
\begin{figure}[tb]
\epsfxsize=7.4cm
\epsffile{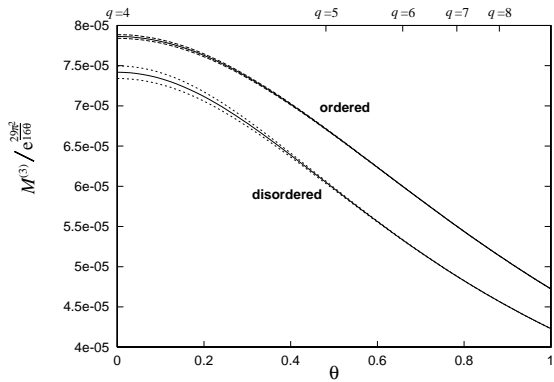}
\caption{
The behavior of the ratio of the absolute value of
the third magnetization cumulants 
$|M_{d,o}^{(3)}|$ to $x^{29/8}$.
         }
\label{fig20}
\end{figure}
\clearpage

\noindent
In conclusion the Pad\'e data of the large-$q$ series 
for the magnetization cumulants also convince us that 
the conjecture (\ref{eq:asymp_m}) for the asymptotic form 
is true at least for $n=2$ and $3$. 

\section{Summary}
The finite lattice method was applied to generate the large-$q$ 
expansion of the $n$th energy cumulants ($n=0,1,\cdots,6$) 
and of the $n$th magnetization cumulants ($n=1,2$ and $3$) 
at the phase transition point for the ordered and disordered phases 
in the two-dimensional $q$-state Potts model.
It enabled us to extend the series of the energy cumulants 
to the 23rd order in $z=1/\sqrt{q}$ for $n=0,1,2$ 
and to the 21st order for $n=3,\cdots,6$ 
from the 10th order generated by Bhattacharya {\em et al.}
using the standard diagrammatic method. 
It also enabled us to generate the new series of the magnetization cumulants
to the 21st order. 
We made the Pad\'e analysis of the series for the product 
of each cumulant and some powers of the latent heat 
that is expected to be a smooth function of $z$ 
if the conjecture by Bhattacharya {\em et al.} is true 
that the relation between the cumulants and the correlation 
length in the asymptotic behavior for $\beta\to\beta_t$ at $q=4$ 
will be kept in their asymptotic behavior 
at the first order transition point for $q\to 4_+$. 
It allowed us to present very precise estimates of the cumulants 
at the transition point for $q>4$.
The accuracy of the estimates is orders of magnitude higher than 
the results of the Monte Carlo simulations and the low-(and high-) 
temperature series for $q\gtsim 7$, and 
for instance of about 0.1 percent at $q=5$ 
where the correlation length is as large 
as a few thousand lattice spacings
and there is no previous data to be compared with, 
and of a few percent even at the limit of $q=4$ 
where the correlation length diverges.
The resulting asymptotic behavior of the cumulants for $q$ close to 4 
is consistent with the conjecture by Bhattacharya {\em et al.}
not only for the second energy cumulant (or the specific heat) 
but also for the higher energy cumulants. 
It is important that the asymptotic behavior of the magnetization cumulants 
is also consistent with the conjecture, which was not investigated by 
Bhattacharya {\em et al.}. This strongly suggests that 
the conjecture would be the case 
for all the physical quantities that diverge or vanish at the first order 
phase transition point for $q\to 4_+$.

\section*{Acknowledgements}
We would like to thank W. Janke, T. Bhattacharya and R. Lacaze 
for valuable discussions.
This work is supported in part 
by the Grants-in-Aid of the Ministry of Education (No.08640494).


\end{document}